\documentclass[aps,pra,superscriptaddress,floatfix,twocolumn]{revtex4-1}
\usepackage{graphicx} 
\usepackage{amsmath}
\usepackage{cancel}
\usepackage{lipsum}
\usepackage{verbatim}
\usepackage{color}
\begin{document}

\title{Dynamics of Quantum Droplets in a Quasi-one-dimensional Framework: An Analytical Approach}
\author{Akshat Pandey}\thanks{Current address: Department of Physics and Astronomy, KU Leuven, Celestijnenlaan 200d, Leuven B-3001, Belgium}
\affiliation{Department of Physics, Shiv Nadar Institution of Eminence Deemed to be University,\protect\\ Greater Noida, Uttar Pradesh-201314, India}
\author{Ayan Khan}\thanks{ayan.khan@bennett.edu.in}
\affiliation{Department of Physics, School of Engineering and Applied Sciences, Bennett
University, Greater Noida, UP-201310, India}



\begin{abstract}
Quantum droplets have been recently observed in dipolar Bose-Einstein condensates (BECs) and in BEC mixtures. This forms the motivation for us to explore the dynamics of these droplets. We make use of the Extended Gross-Pitaevski equation which apart from the effective mean field (MF) interaction, also includes a \textit{beyond} mean field interaction. The competition of these two interactions in the context of droplet formation is explored. Further, the conditions for droplet formation are studied. 
\end{abstract}
\maketitle
\section{Introduction}
 The study of particle behaviour at ultra-low temperatures has been of continual interest \cite{tanzi, kadau, li, englezos} ever since Bose-Einstein condensates (BECs) were first experimentally obtained \cite{dav}. The curiosity in the physics of BEC has been of due to its application in varied range of topics including dark matter phenomenology \cite{dm}, analogue gravity models \cite{visser}, simulating condensed matter theories \cite{roati,hart} and Quantum information processing \cite{qip}.

The recent experimental observation of a liquid-like state in BEC mixtures \cite{cab} and dipolar BEC \cite{pfau} has opened new domains of research in the understanding of the liquid state which truly manifests a quantum many-body state as compared to the conventional liquid. This is particularly interesting as our usual understanding of the liquid state is based on the theory of Van der Waals (or some extension thereof) \cite{ifb}. According to this theory, the liquid state is a consequence of an equilibrium between the attractive inter-particle forces and the repulsive short range forces, which is obtained at high densities \cite{cap}. However, this new liquid-like state is peculiar as its properties do not conform to those of the classical liquids. This is so because in ultra-cold atoms, the inter-particle spacing is larger as compared to Van der Waals liquids and the densities are relatively low. Instead, this liquid-like state is described by the beyond mean field (BMF) Extended Gross Pitaevskii (EGP) theory, which is obtained by employing Lee-Huang-Yang (LHY) \cite{lhy} corrections to the traditional mean field (MF) Gross Pitaevskii (GP) theory. The GP equation has the same form as the Schr\"{o}dinger equation with an additional non-linear (cubic) term, while the EGP equation has a supplementary quartic term. However, the interpretation of these relatively similar equations is much different, while the Schr\"{o}dinger equation describes the dynamics of a one-particle wave-function, GP equation and EGP equation describe the dynamics of a collection of particles or the matter waves. The existence of competing interactions in the EGP equation further allows us to study the liquid like properties yielding from the condensate. The competition of MF and BMF interactions can lead to negative effective interaction which results in the clustering of atoms where constant density can be observed over the spatial domain. This is the onset of liquid formation which we define as quantum droplets \cite{ifb}.

BECs are experimentally obtained in a three-dimensional magneto-optic trap, which is usually isotropic and harmonic in nature. However, we can relax the isotropic condition reducing the number of effective degrees of freedom via effective tuning of the external counter propagating laser sources. This trap engineering can introduce strong confinements in the transverse direction, and reduce the system to effectively one dimension or vice-versa. When we have strong confinement along the transverse direction, the condensate is allowed to expand in the longitudinal direction thereby describing an effective one dimensional system which is commonly known as quasi-one-dimension. The  Quasi-one-dimensional (Q1D) geometry is viable experimentally, as in principle it is still a three dimensional geometry. Therefore, unlike a true one-dimensional (1D) geometry, a Q1D geometry can allow the formation of condensate \cite{petrov}. 

The modifications to the GP theory have different manifestations in 1D and 3D(and thus Q1D). The dynamics of quantum droplets (QDs) have been extensively studied in 1D, see, for example \cite{malomed1, malomed2, nath,zez, abbas} where the beyond mean field interaction is manifested via quadratic non-linearity.  However, the cubic-quartic non-linearity is relatively less explored (in Q1D).  For this reason and for reasons related to experimental amenability, in this paper, we study the dynamics of quantum droplets in Q1D geometry analytically. 

 The present paper is organized as follows. In Sec. \ref{theory}, we describe the theoretical framework where we reduce the system from 3+1 dimension to 1+1 dimension which gives us an equation governing BECs in Q1D. Later in Sec. \ref{analysis}, we present the analytical solutions and analyze their physical implications. We look at the conditions for droplet formation and droplet-soliton crossover. We draw our conclusions in Sec. \ref{conclusion}. 
 
\section{Theoretical Framework}\label{theory}
Quantum liquid or droplet in ultra-cold atomic gases received considerable attention through the seminal work of Petrov, where he had suggested a stabilization mechanics via beyond mean-field interactions in a binary BEC \cite{petrov} leading to the clustering of atoms to form droplet like state. It is interesting to note that, similar clustering mechanism was predicted two decades back by Bulgac where the competition between two body mean-field interaction and three body interaction was predicted to play the pivotal roles \cite{bulgac}. However, recent experiments supported the beyond mean-field mechanism while the clustering due to the three body effect is yet to be verified and considered as challenging by experimentalists \cite{barbut_PT}.

We work with a two component Bosonic mixture, which essentially is distributed in just one dimension, due to the presence of a cigar shaped trap. Usually in a binary BEC mixture, one can identify two distinct states (miscible and immiscible), corresponding to the interplay of inter-species and intra-species interactions. In case of a strong attractive intra-species interaction, the mixture can collapse from a miscible phase. If we consider the two components to have the same spatial mode, we can neglect the spin excitations in the atoms i.e we use the mean-field approximation. In this case, the mixture, close to the collapse point in the miscible phase, can be described by a single component $\Psi$.  The dynamics of $\Psi$ can then be described by the Extended Gross Pitaevskii (EGP) equation \cite{rev}. In this section, we start from the EGP equation in 3+1 dimension and employ a dimension reduction scheme to reduce the system dimension to 1+1 \cite{deb}. One essential difference from our previous investigations is that, here the trap modulation and change in interaction strength over time is taken care of. The governing EGP can be noted as, 
 \begin{equation}
i \hbar \frac{\partial \Psi}{\partial t}=\left[\left(-\frac{\hbar^{2}}{2 m} \nabla^{2}+V_{trap}\right)+U|\Psi|^{2}+U^{\prime}|\Psi|^{3}\right] \Psi.\label{egp_3d}
\end{equation}
Here, $U$ and $U'$ denotes the effective two-body meanfield interaction and beyond meanfield interaction respectively. Quantitatively, $U=\frac{4\pi\hbar}{m}\delta a$ and $U'=\frac{256\sqrt{\pi}\hbar^2\delta a'}{15m}$, where $m$ being the mass of the atoms. $\delta a$ and $\delta a'$ relates to the intra-species ($a_{11}$ and $a_{22}$) and inter-species ($a_{12}$) $s$-wave scattering lengths respectively such that $\delta a\propto a_{12}+\sqrt{a_{11}a_{22}}$ and $\delta a'\propto(\sqrt{a_{11}a_{22}})^{5/2}$.

In order to theoretically model the Q1D geometry, we employ the following ansatz \cite{khan},
    \begin{equation}
\Psi(\mathbf{r}, t)=\frac{1}{\sqrt{2 \pi a_{B}} a_{\perp}} \psi\left(\frac{x}{a_{\perp}}, \omega_{\perp} t\right) e^{\left(-i \omega_{\perp} t-\frac{y^{2}+z^{2}}{2 a_{\perp}^2}\right)}.\label{ansatz}
\end{equation}
Here $\omega_{\perp}$ is the transverse trapping frequency which is typically 10 times larger than $\omega_{0}$, the longitudinal frequency.  This accounts for the strong confinement in the transverse directions. This high value for the potential trap in the transverse directions leads to its cigar-like shape. Here $a_{\perp}=\sqrt{\frac{\hbar}{m \omega_{\perp}}}$ is the characteristic length scale corresponding to $\omega_{\perp}$ and $a_{B}$ is the Bohr radius. We also note that, from here onward, we will use the dimensionless notation for the spatio-temporal variables i.e. $x \equiv x/a_{\perp}$ and $t \equiv \omega_{\perp}t$.\\

Plugging Eq.(\ref{ansatz}) into Eq.(\ref{egp_3d}) we end up with a Q1D EGP equation for $\psi$ such that,
\begin{eqnarray}
    i \frac{\partial \psi(x, t)}{\partial t}&=&\left[-\frac{1}{2} \frac{\partial^{2}}{\partial x^{2}}+\frac{1}{2} M(t) x^{2}-g_{1}(t)|\psi(x, t)|^{2}\right.\nonumber\\&&\left.+g_{2}(t)|\psi(x, t)|^{3}\right] \psi(x, t)\label{egp_1d}
\end{eqnarray}

In Eq.(\ref{egp_1d}), $g_{1}=-2\delta a/a_{B}$, $g_{2}=(64\sqrt{2}/15\pi)\delta a^{\prime} /\left(a_{B}^{3 / 2} a_{\perp}\right)$ and $M=\omega_{0}^{2}/\omega_{\perp}^{2}$.
Note, here in the case of the Q1D geometry, an effective negative (attractive) MF interaction is counterbalanced by a positive (repulsive) BMF interaction. This is unlike the 1D case, where the MF and BMF interactions were reversed i.e., repulsive and attractive respectively. Note also that, the BMF term in this geometry is quartic in $\psi$ while in the case of a purely 1D geometry, the BMF term is quadratic in $\psi$ \cite{petrov_1d}. In the following section, we analyze Eq.(\ref{egp_1d}) which now carries distributed coefficients. The time varying nature of the harmonic confinement and the interaction have deep implications towards the pursuit of coherent control.

\section{Solutions}\label{analysis}
To study the dynamical behaviour of the condensate analytically, it is convenient to transfer Eq.(\ref{egp_1d}) in the centre of mass frame.
In order to do this, we make use of the following ansatz, prescribed in \cite{nath}
 \begin{equation}
    \psi(x, t)=\sqrt{A(t)} F[\xi(x, t)] e^{i \theta(x, t)}
\end{equation}
Here $A(t)$ is the time modulated amplitude. $\xi(x,t)$ is the travelling coordinate of the condensate, such that $\xi=\alpha(t) x+\eta$. Here, $\alpha$ is positive definite function of time with $\eta>0$. The non trivial phase factor $\theta (x,t)$ is defined as, 
$$
\theta(x, t)=-\frac{\alpha^{\prime}(t)}{2 \alpha(t)} x^{2}-\int \frac{\epsilon \alpha^{2}(t)}{2} \partial t
$$
Here, $\alpha(t)$ corresponds to the inverse of the width of the excitation. $\alpha^{\prime}(t)$ is the first order time derivative of $\alpha(t)$
Plugging this ansatz for $\psi$ into Eq.(\ref{egp_1d}), it is possible to separate the equation into two by collecting the real and imaginary parts of the equation and equating them to zero. 
The imaginary equation can be noted as,
\begin{eqnarray}
&&\frac{1}{2 \sqrt{A}} \frac{\partial A}{\partial t} F e^{i \theta}+\sqrt{A} \frac{d F}{d \xi} \frac{\partial \xi}{\partial t} e^{i \theta}\nonumber\\&&+\sqrt{A} \frac{d F}{d \xi} \frac{\partial \xi}{\partial x} \frac{\partial \theta}{\partial x} e^{i \theta}+\frac{\sqrt{A}}{2} F e^{i \theta} \frac{\partial^{2} \theta}{\partial x^{2}}=0\label{egp_im_1}
\end{eqnarray}
Upon solving this equation, we end up with
\begin{eqnarray}
    \frac{1}{2 A} \frac{d A}{d t} F&+&\frac{F}{2}\left(\frac{-\alpha^{\prime}(t)}{\alpha(t)}\right)=0\nonumber\\
\textrm{or,}\,\, A(t)&=&\alpha(t),\label{egp_im_2}
\end{eqnarray}
which result implies the coupling of phase with amplitude. Here, we have used the fact that, $\frac{\partial\theta}{\partial x}=-\frac{\alpha'}{\alpha}x$, $\frac{\partial^2\theta}{\partial x^2}=-\frac{\alpha'}{\alpha}$, $\frac{\partial\theta}{\partial t}=\frac{-\alpha\alpha''+\alpha'^2}{2\alpha^2}x^2-\frac{\epsilon\alpha^2}{2}$, $\frac{\partial\xi}{\partial t}=\alpha'(t)x$, $\frac{\partial\xi}{\partial x}=\alpha(t)$ and $\frac{\partial^2\xi}{\partial x^2}=0$.

The real part of Eq.(\ref{egp_1d}) can be described as,
\begin{eqnarray}
&&-\sqrt{A} F \frac{\partial \theta}{\partial t} e^{i \theta}+\frac{1}{2} \sqrt{A} \frac{d^{2} F}{d \xi^{2}}\left(\frac{\partial \xi}{\partial x}\right)^{2} e^{i \theta}\nonumber\\&&-\frac{\sqrt{A}}{2} F e^{i \theta}\left(\frac{\partial \theta}{\partial x}\right)^{2}
 -g_{1}(t)(\sqrt{A} F)^{3} e^{i \theta}\nonumber\\&&+g_{2}(t)(\sqrt{A} F)^{4} e^{i \theta}-\frac{1}{2} M(t) x^{2} \sqrt{A} F e^{i \theta}=0\label{egp_re_1}
\end{eqnarray}
Using the consistency condition derived in Eq. (\ref{egp_im_2}) and carrying out coordinate transformation to the centre of mass frame, we get
\begin{eqnarray}
&&\frac{\sqrt{\alpha} F \epsilon \alpha^{2}(t)}{2}-\sqrt{\alpha(t)} F\left(\frac{2 \alpha^{\prime 2}(t)-\alpha^{\prime \prime}(t) \alpha(t)}{2 \alpha^{2}(t)}\right) x^{2}\nonumber\\&&+\frac{\sqrt{\alpha(t)}}{2} \frac{d^{2} F}{d \xi^{2}} \alpha^{2}-g_{1}(t)(\sqrt{\alpha(t)} F)^{3}\nonumber\\&&+g_{2}(t)(\sqrt{\alpha(t)} \xi)^{4}
-\frac{1}{2} M(t) x^{2} \sqrt{\alpha(t)} F=0.\label{egp_re_2}
\end{eqnarray}
$\alpha^{\prime\prime}(t)$ is the second time derivative of $\alpha(t)$.\\
This gives us another consistency condition
\begin{equation}
   M(t)=\frac{\alpha^{\prime \prime}(t) \alpha(t)-2 \alpha^{\prime^{2}}(t)}{2 \alpha^{2}(t)}.\label{trap1}
\end{equation}
It is interesting to note here that by applying a transformation $\alpha(t)=1/\beta(t)$, Eq.(\ref{trap1}) reduces to 
\begin{eqnarray}
\beta''(t)+2M(t)\beta(t)=0.\label{trap2}
\end{eqnarray} 
This implies that the current analytical scheme actually couples the amplitude of the wavefunction with the trapping configuration. 
This can further be mapped to the well known Riccati type equation, $\sigma'(t)-\sigma^2(t)=2M(t)$, if we consider $\beta(t)=e^{-\int_0^t\sigma(t')dt'}$.
Eq.(\ref{trap2}) can lead to sinusoidal behaviour of $\beta$ for the most trivial case, i.e., for regular harmonic oscillators (when $M$ is positive and constant).
For negative and constant $M$, one can experience an expulsive oscillator. The situation can become more curious for time modulated confinements. 
The periodic variation can lead to a Mathieu function solution which has direct implications in quantum pendulums.
However, the nature of $M\equiv M(t)$ can not be only restricted to oscillatory behaviour and thus can have a host of solutions with different physical implications \cite{nath1}.
 
Applying Eq.(\ref{trap1}) in Eq.(\ref{egp_re_2}) we obtain,
\begin{equation}
    \frac{\alpha^{5 / 2}(t)}{2} F \epsilon+\frac{\alpha^{5 / 2}(t)}{2} \frac{\partial^{2} F}{\partial \xi^{2}}-g_{1}(t) \alpha^{2}(t) F^{3}+g_{2}(t) \alpha^{3 / 2}(t) F^{4}=0.\label{egp_re_3}
\end{equation}

In order to cast Eq.(\ref{egp_re_3}) into a more convenient form, we define two non-linearity constants $G_1$ and $G_2$ such that 
\begin{equation}
    g_{1}(t)=\frac{G_{1} \alpha^{1 / 2}(t)}{2};\qquad  g_{2}(t)=\frac{G_{2} \alpha(t)}{2},\label{1g1g2}
\end{equation}
as a result, Eq.(\ref{egp_re_3}) can now be written as,
\begin{equation}
    -\frac{d^{2} F}{d \xi^{2}}+G_{1}|F|^{2} F-G_{2}|F|^{3} F=\epsilon F. \label{egp_re_4}
\end{equation}
Here, $\epsilon$ is the eigenvalue. The current analytical treatment allows us to consider a more generic form of the dynamical equation and it automatically takes into account temporal variation for interaction along with the trap (which we discussed earlier). Since the interaction is controlled by an external magnetic field through Feshbach resonance, therefore a temporal variation of the interaction is always physically meaningful and the current analytical scheme captures this possibility seamlessly.

It is important now to recognize that Eq.(\ref{egp_re_4}) is already mentioned in Ref.~\cite{deb}. Hence, we can now pick up the stable solution described there to study the dynamical behaviour of the droplets. Applying the prescribed ansatz solution, \begin{equation}
    F(\xi)=\frac{P}{1+\sqrt{1-P} \cosh (\sqrt{\lambda} \xi)},
\end{equation}
we obtain a set of constrained condition such that, 
\begin{equation}
    \begin{aligned}
&P=\frac{\lambda \pm \sqrt{3 \epsilon G_{2}-3 G_{2} \lambda+\lambda^{2}}}{G_{1}}\\
&\lambda=3 \epsilon-G_{1} \\
&\epsilon=G_{1}-G_{2} \\
&\left|G_{1}\right|=2 G_{2} \text { or }\left|G_{1}\right|=3 G_{2}
\end{aligned}
\end{equation}
The last of these implies that analytical solutions are only possible for two specific ratios of $G_1$ and $G_2$ that is, for specific ratios of MF and BMF interaction strengths.
In terms of these constraint conditions, the solutions for $F(\xi)$ turn out to be
\begin{equation}
    \begin{aligned} F(\xi) &=\frac{-12 \mu_{G_{1}}}{1+\sqrt{1+12 \mu_{G_{1}}} \cosh \left(\sqrt{\frac{-G_{1}}{2}} \xi\right)} \text { for }\left|G_{1}\right|=2 G_{2} \end{aligned}
\end{equation}
\begin{equation}
    \begin{aligned} =\frac{1-12 \mu_{G_{1}}}{1+\sqrt{-12 \mu_{G_{1}}} \cosh \left(\sqrt{-G_{1}} \xi\right)} \text { for }\left|G_{1}\right|=3 G_{2} \end{aligned}
\end{equation}
Here $\mu_{G{1}}=\frac{\mu_{0}}{G_{1}}$ where $\mu_{0}$ is the chemical potential.
Therefore, this can be put back into the ansatz to yield the complete set of solutions for $\psi$
\begin{eqnarray}
   &&  \psi(x, t)=\frac{\left(-12 \mu_{G_{1}}\right)\sqrt{\alpha(t)}}{1+\sqrt{1+12 \mu_{G_{1}}} \cosh \left(\sqrt{\frac{-G_{1}}{2}}\left( \alpha(t) x+\eta\right)\right)}\nonumber\\&&\times e^{i\left(-\frac{\alpha^{\prime}(t)}{2 \alpha(t)} x^{2}-\int \frac{\epsilon \alpha^{2}(t)}{2} \partial t\right)} \text { for }\left|G_{1}\right|=2 G_{2}\label{sol1}
\end{eqnarray}

\begin{eqnarray}
   &&  \psi(x, t)=\frac{\left(1-12 \mu_{G_{1}}\right)\sqrt{\alpha(t)}}{1+\sqrt{-12 \mu_{G_{1}}} \cosh \left(\sqrt{-G_{1}}\left( \alpha(t) x+\eta\right)\right)}\nonumber\\&& e^{i\left(-\frac{\alpha^{\prime}(t)}{2 \alpha(t)} x^{2}-\int \frac{\epsilon \alpha^{2}(t)}{2} \partial t\right)} \text { for }\left|G_{1}\right|=3 G_{2}\label{sol2}
\end{eqnarray}
We have noted that among the two solutions proposed, one solution emerged as stable and thus we make use of that solution here for further analysis. The stable solution corresponds to 
$|G_1|=3G_2$ \cite{deb}. Hence, we will focus on Eq.(\ref{sol2}) from here on.

We are now interested in examining the dynamical behaviour of Eq.(\ref{sol2}) for two different potential landscapes. First we will consider the homogeneous system for which $M(t)=0$ and secondly we will take into account the regular harmonic potential where $M(t)=M^2$.

\subsection*{Quantum Droplets in a homogeneous system $(M(t)=0)$}
Here we analyze the solution in the absence of an external confinement. Within our model, the most trivial way to incorporate $M(t)=0$ is by considering $\alpha(t)$ is constant (see Eq.(\ref{trap1})). This will a yield static solution.  Physically, this is equivalent to the condition that the transverse frequency is much larger than the longitudinal frequency $(\omega_0\ll\omega_{\perp})$, implying $M\xrightarrow{}0$. 

However, from Eq.(\ref{trap2}) we can also infer that $\alpha(t)=\frac{1}{C_1+C_2t}$, where $C_1$ and $C_2$ are integration constants if we solve Eq.(\ref{trap2}) and this solution also satisfies Eq.(\ref{trap1}). This will provide dynamics of localized or droplet density profiles based on the particle number.

\begin{figure}[!h]
\centering
\includegraphics[scale=0.35]{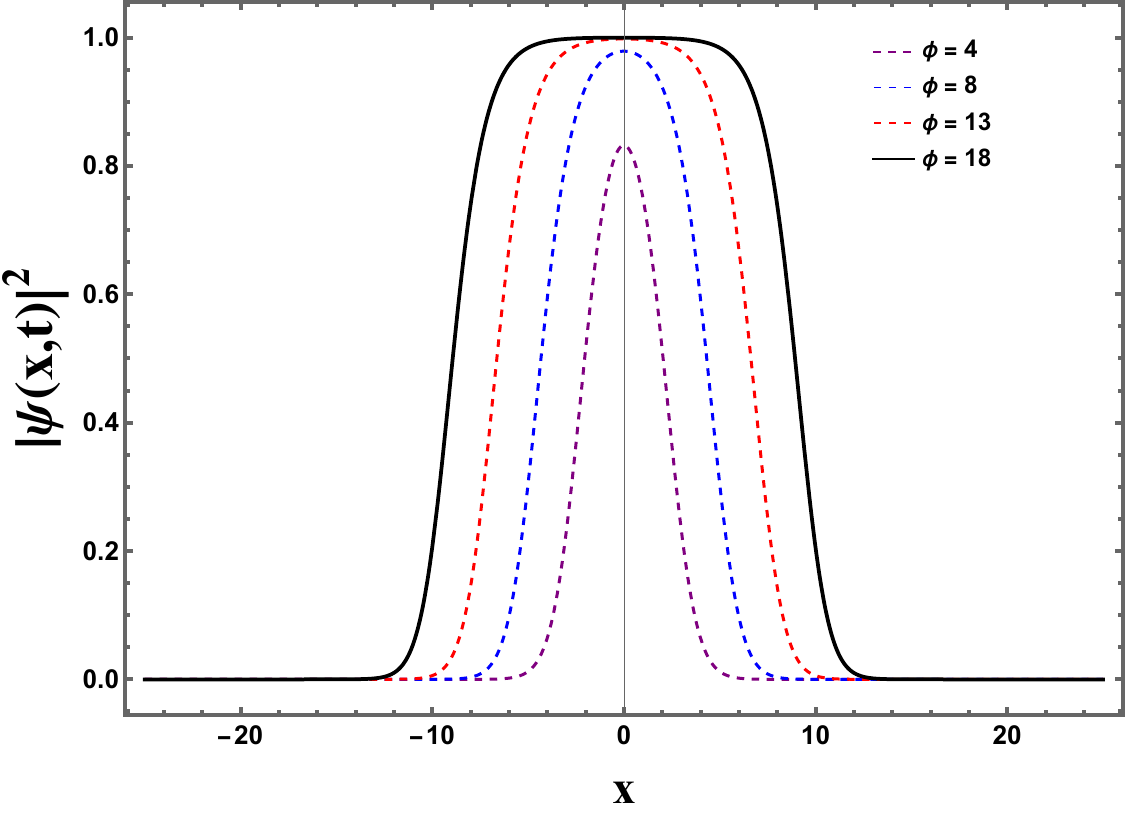}
\caption{(Colour Online) The static density profiles in free space, parameterized by the particle number following Eq.(\ref{pnumber})$\phi$, where $\alpha_0 = 1.$, $G_1 = -1$. All parameters are in arbitrary units. The figure depicts the transition of localized density distribution into a flat-top one with higher particle number. The solid black line corresponds to $\phi=18$, while the dashed red, blue and purple line depicts $\phi=13, 8\,\textrm{and}\,4$ respectively.}
\label{f1}
\end{figure}

Note also that since there is no external confinement, all the non-triviality in the dynamics, including the formation of droplets and solitions, is due to the cubic-quartic interactions. Therefore the goal here is to understand the interplay between these MF and BMF interactions and the role of the chemical potential in droplet/soliton formation.
For this purpose and for computational simplicity, we consider, $\alpha>0$, $\mu_{G_{1}}>0$ and $\eta$ is set to $0$. In Fig.~\ref{f1} the condensate density profiles for different particle numbers have been plotted where the particle number $\phi$ is defined as
\begin{equation}
\phi=\int_{-\infty}^{\infty}|\psi(x)|^{2} d x\label{pnumber}
\end{equation}

We have considered decreasing values of the chemical potential $(\mu_0)$ which correspond to increasing values of the particle density. We see from the figure that with increasing particle number, we get flatter tops in the density profiles i.e., constant density over a finite region in space. This suggests the clustering of atoms where its density remains unchanged over a certain spatial degree. This plateau formation is considered as the signature of droplet formation. Low particle number gives sharper peaks, in the density profiles i.e., localized modes, which correspond to the solitonic regimes.
However, these observations are nothing new and have been reported in the recent past \cite{deb}. Nevertheless, what makes this study unique, is incorporation of the dynamical behaviour through an analytical scheme for a quasi one dimensional system.

\subsection*{Quantum Droplets in a regular harmonic trap $(M(t)=M^{2})$}
Since BECs are obtained in traps, it is rather useful to explore the case for a regular harmonic trap potential. For regular harmonic trap (say, $M(t)=M^2$), we can easily deduce from Eq.(\ref{trap2}) that $\beta(t)=\beta_0\cos{\sqrt{2}Mt}$ resulting $\alpha(t)=\alpha_{0}\sec(\sqrt{2}Mt)$ with $\alpha_{0}>0$.
Setting $\eta=0$, the resultant form of the complete wavefunction then becomes
\begin{eqnarray}
 &&    \psi(x, t)=\frac{\left(1-12 \mu_{G_{1}}\right)\sqrt{\alpha_{0}\sec(\sqrt{2}Mt)}}{1+\sqrt{-12 \mu_{G_{1}}} \cosh \left(\sqrt{-G_{1}}\left( \alpha_{0}\sec(\sqrt{2}Mt) x\right)\right)}\nonumber\\&&\times e^{i\left(-\sqrt{2}M \tan(\sqrt{2}Mt) x^{2}-\int \frac{\epsilon \alpha_{0}^{2}\sec(\sqrt{2}Mt)}{2}  dt\right)} \label{trap_d}
\end{eqnarray}

\begin{figure}[!h]
\centering
\includegraphics[scale=0.25]{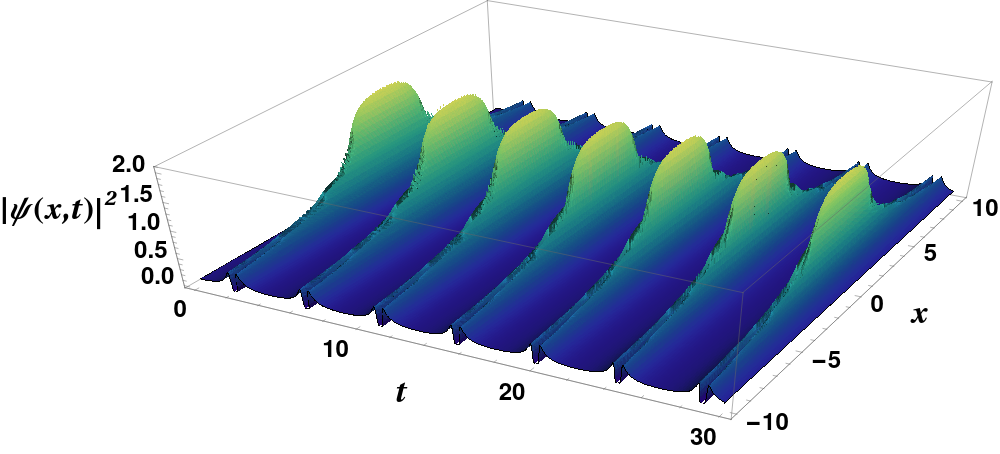}
\caption{The condensate density profiles displayed with varying time with $M=0.5$, $\alpha_0=0.08$, $\mu_{G_1}=0.000001$ and $G_1=-1$. The parameter values are in arbitrary units.}
\label{f2}
\end{figure}

In Fig.~\ref{f2} we have plotted the temporal dynamics obtained from Eq.(\ref{trap_d}). Here $M$ is set at $0.5$. Since the temporal evolution of the density profile has explicit dependence on the secant function, thus we observe rapid variation in the vicinity $t=n\pi/2$. Physically, this corresponds to the melding of the condensate atoms leading to a compression of condensate atomic density, due to temporal modulation of the interaction. This also explicates the collapse and revival of the droplets over time evolution due to the temporal variation of the interaction. However, the intertwining nature of interaction and trapping with time explicated through this model demands a deeper insight on their role in time dynamics.

\subsection*{Role of Interaction and Trapping}
\begin{figure}[!h]
\centering
\includegraphics[scale=0.35]{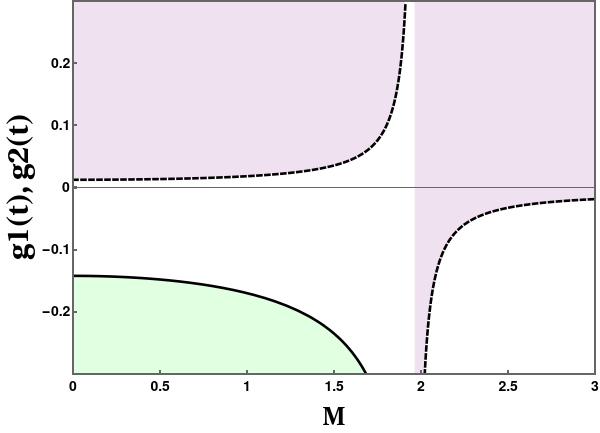}
\caption{(Colour Online) Variation of MF and BMF interaction with respect to the modulating trap is depicted. The black dashed line and solid line draw the contours of BMF and MF interactions respectively. The figure is snapped at $t=0.8$. The other parameters are $\alpha_0 = 0.08$, $G_1=-1$ and $G_2 = 0.33$.}
\label{f3}
\end{figure}

Firstly, we examine the competition between mean-field and beyond mean-field interaction as a function of trapping strength. For that purpose, we plot $g_{1}(t)$ and $g_{2}(t)$ against $M$ at constant time as depicted in Fig.~\ref{f3}. For convenience, we also note down the explicit functional dependencies in Eq.(\ref{g1g2})
\begin{equation}   
g_{1}(t)=\frac{G_{1} \left(\alpha_{0}sec(Mt)\right)^{1 / 2}}{2};\quad g_{2}(t)=\frac{G_{2} \alpha_{0}sec(Mt)}{2}.\label{g1g2}
\end{equation}

The presence of the $\sec$ function in Eq.(\ref{g1g2}) it is inevitable that there exists singular points. We observe that, the presence of attractive mean-field interaction (light green shaded region bordered by solid black line) and a repulsive beyond mean field interaction (light purple shaded region with dashed black line) in the region between $M=0$ and $M=\pi/2$. Since the attractive term is proportional to $\sqrt{\alpha(t)}$, we require $\alpha_0>0$ in order to keep the MF interaction real and we set $\alpha_0=0.08$. Then, the competition of the attractive and repulsive interactions leads to the formation of quantum droplets in this region of $M$. Similar trends can be observed for $3\pi/2<M<2\pi$ and so on. For the region between $M=\pi/2$ and $M=3\pi/2$, the sec function and consequently $\alpha(t)$ has negative values. Thus, the mean field term, which is proportional to $\sqrt{\alpha(t)}$ turns imaginary and thus the only real interaction (in this region of $M$) is the repulsive beyond mean field interaction which leads to the formation of solitons. This analytical model also suggests that, it is possible to observe a reverse scenario for $\alpha_0<0$ where droplet formation can be possible in the region $3\pi/2<M<3\pi/2$. Thus, in the purview of the current analytical scheme, it points toward the possible dropleton-soliton transition in the region $M=n\pi/2$. 

\begin{figure}[!h]
\centering
\includegraphics[scale=0.25]{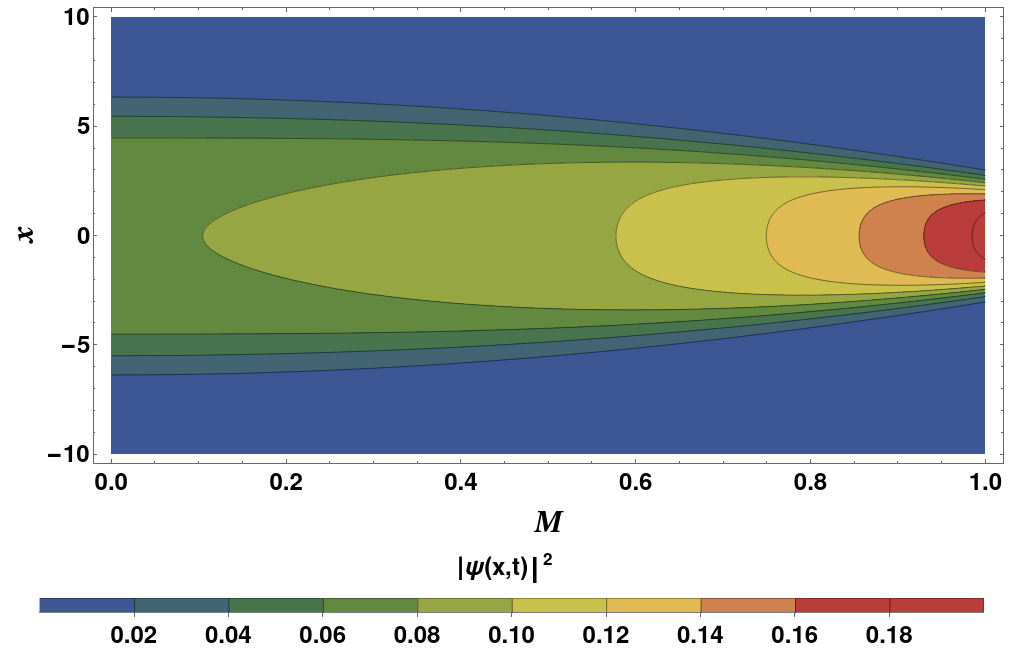}
\caption{Variation of external trapping frequency resulting in transition from flat top to localized structure. The time freezed snapshot is captured as $t=0.8$ with $\alpha_0=0.08$, $\mu_{G_1}=0.000001$ and $G_1=-1$.}
\label{trap}
\end{figure}

Here, it must be noted that in one of our recent investigations, we have shown numerically that weaker longitudinal frequency supports flat-top structure while as we tighten the longitudinal frequency we observe a transition toward localized solution which is correctly captured in the current analytical model where droplet formation is supported in the region $0<M<\pi/2$. However, the transition is fairly smooth and lacks any explicit critical point \cite{debnath}. In Fig.~\ref{trap} we have tried to replicate the same through our analytical model and we find our analytical model is good enough to capture the same picture.

However, the additional suggestion of collapse and revival of droplet formation at higher trapping potential is subject to future verification. Nevertheless, a similar prediction can also be found in the 1D quantum droplet in Ref.\cite{nath}. It is also noteworthy that several interesting results have recently emerged by subtle changes to the trap. For example, by adding a temporal perturbation to the harmonic trap \cite{nath2}, by using a double-well potential \cite{doublewell}, or in the presence of narrow wells and barriers \cite{debnath2}. Additionally, apart from the QD, dynamics of the solitonic phase have been studied in detail \cite{debnath3, gangwar}. Another interesting paper is by Katsimiga \textit{et. al} which amongst other things explores the dynamics of bubbles, and of limiting cases of QDs, namely stable kinks \cite{malomed3}.

\section{Conclusions}\label{conclusion}
In this work, we studied the dynamics of droplets in a Q1D two-component BEC mixture. Assuming the two components occupy the same spatial mode \cite{che}, we got to an effectively one-component EGP equation in three dimensions. Further, we carried a dimensional reduction mechanism to obtain the EGP equation in Q1D. We note that this equation is different from its 1D counterpart, particularly in the nature of the interactions, and the order of exponent in the BMF term. This work, thus, also explored the relatively unexplored quartic non-linearity in the EGP equation. An appropriate scaling of the dynamical equation allows us to use recently obtained analytical solutions in Q1D geometry. However, the current system reveals the intertwining nature of the interaction and trapping potential and their explicit time modulation. Hence, in the later part we investigate their role in soliton-dropleton transition by studying their explicit time dependence. The current analytical model suggests that it is possible to control the interaction through external trapping frequency and it is also possible to observe the soliton-dropleton transition by modulating the trap. These observations are fully consistent with the predictions made for 1D Bose gas \cite{nath}. However, it will be interesting to explore the experimental feasibility.
 
\section*{Acknowledgement}
AK thanks the Council of Scientific and
Industrial Research (CSIR) Human Resource
Development Group (HRDG) Extramural Research Division (EMR-II), India for the
support provided through project number
03/1500/23/EMR-II.

\section*{Data Availability}
Data will be made available on request.

\section*{Author Contribution}
AP carried out the calculation. AK conceived the work. Both authors contributed equally in analysis and writing.
\bibliographystyle{apsrev4-1}
\bibliography{refs}

\end{document}